\documentclass[5p,twocolumn]{elsarticle}
\usepackage{amsmath,amssymb,amsfonts,bbm,graphicx,times,color,mathptmx,hyperref}
\usepackage{mathrsfs}
\journal{Optics Communication}
\newcommand{\Tr}{{\rm Tr}}
\newcommand{\iid}{\mathbb{I}}
\newcommand{\bmsigma}{\boldsymbol \sigma}
\newcommand{\bmLambda}{\boldsymbol \Lambda}
\newcommand{\bmX}{\boldsymbol X}
\newcommand{\bmR}{\boldsymbol R}
\newcommand{\sfx}{{\sf x}}
\newcommand{\sfy}{{\sf y}}
\newcommand{\ha}{\hat{a}}
\newcommand{\hb}{\hat{b}}
\newcommand{\hk}{\hat{k}}
\newcommand{\hx}{\hat{x}}
\newcommand{\hq}{\hat{q}}
\newcommand{\hp}{\hat{p}}
\newcommand{\hN}{\hat{N}}
\begin{document}
\begin{frontmatter}
\title{Homodyning the $g^{(2)}(0)$ of  Gaussian states}
\author{Stefano Olivares}
\ead{stefano.olivares@fisica.unimi.it}
\author{Simone Cialdi}
\ead{simone.cialdi@mi.infn.it}
\author{Matteo G. A. Paris}
\ead{matteo.paris@fisica.unimi.it}
\address{Quantum Technology Lab, Dipartimento di Fisica {``Aldo Pontremoli''}, 
Universit\`a degli Studi di Milano, I-20133 Milano, Italy. \\
Istituto Nazionale di Fisica Nucleare, Sezione di Milano, 
I-20133 Milan, Italy.}
\begin{abstract}
We suggest a method to reconstruct the 
zero-delay-time second-order correlation function 
$g^{(2)}(0)$ of Gaussian states using a single
homodyne detector. To this purpose, we have found an
analytic expression of $g^{(2)}(0)$ for single- and two-mode
Gaussian states in terms of the elements of 
their covariance matrix and the displacement amplitude.
In the single-mode case we demonstrate our 
scheme experimentally, and also show that when
the input state is nonclassical, there exist a threshold 
value of the coherent amplitude, and a range of values of the
complex squeezing parameter, above which $g^{(2)}(0) < 1$. 
For amplitude squeezing and real coherent amplitude, the 
threshold turns out to be a necessary and sufficient 
condition for the nonclassicality of the state. 
Analogous results hold also for two-mode 
squeezed thermal states.
\end{abstract}
\date{\today}
\end{frontmatter}
\section{Introduction}\label{s:intro}
Gaussian states, namely, states with a Gaussian Wigner 
functions, are fully characterised by the first two moments
of the canonical variables. As a consequence, the full information 
about their quantum state is conveyed by the vector of the 
average values and by their covariance matrix \cite{oli:rev}. 
This class of states, which includes  coherent,
squeezed and two-mode squeezed states, plays a leading role in 
continuous-variable quantum information processing 
\cite{weed:12,ades:14,FOP:05} and high-precision
sensing \cite{rnc:oli}.
\par
The main tool for the experimental characterisation of 
Gaussian states is homodyne detection \cite{h0,h1,h2,h3,h4,h5,h6}, 
which allows one to detect a fixed field-quadrature on the 
input state. The set of data obtained by measuring the quadratures 
at different phase may be then exploited for the tomographic 
reconstruction the quantum state, i.e. the reconstruction of 
the elements of the density matrix, or any other quantity, 
including those not corresponding to a feasible detection 
scheme \cite{dari:03,espo:14,buono:10}. For Gaussian state, it 
is thus natural to seek analytic expressions for any relevant 
quantity in terms on the first and second moments \cite{sera:04},
which, in turn, may be reliably reconstructed by quantum 
tomography, e.g. by sampling the corresponding kernel 
pattern-function or by maximum-likelihood methods \cite{hra04,park14}.
\par
In this paper we focus attention on the zero-delay time 
second-order correlation function which, for a single-mode 
field, may be expressed as
\begin{align}
g^{(2)}(0) = \frac{\langle (\ha^{\dag} \ha)^2 \rangle 
- \langle \ha^{\dag} \ha \rangle}{\langle \ha^{\dag} 
\ha \rangle^2}\,, \label{g2def}
\end{align}
The $g^{(2)}(0)$ correlation function was originally introduced 
to discriminate non-classical anti-bunched light from 
classical thermal one, and currently it continues to play a major
role in the characterisation of  a light source \cite{Roy06}. 
In Eq.~(\ref{g2def})
$\ha$ and $\ha^\dag$ denote the field operators, $[\ha,\ha^\dag] 
= {\mathbbm I}$,
and  $\langle \cdots  \rangle = {\rm Tr}[\varrho \cdots]$, 
$\varrho$ being the density operator 
describing the quantum state of the single-mode field.
\par
In particular, we address single-mode Gaussian states 
and obtain an analytic expression of $g^{(2)}(0)$ for a 
generic Gaussian state in terms of the elements
of its covariance matrix and of the displacement 
amplitude. Similar problems has been addressed before 
\cite{alex:16}, but our approach has a clear practical
advantage, since the covariance matrix and the first moment 
vector are quantities that can be accessed by means of a 
{\it single homodyne detector} \cite{dauria:09}, thus reducing 
the level of complexity of the setups, i.e. 
without the need of full  tomography, photon-resolving 
detectors or double-homodyne detectors \cite{qi:17,teo17}.
\par
In principle, second-order correlation functions may be 
obtained from the photon-number statistics of the input 
state \cite{high:order,bina:QM}. However, the photon-number distribution may not 
be directly accessible, i.e. measuring the distribution 
experimentally may be challenging, or even impossible, 
in several situations. This is the case,
for instance, of continuous-variable states obtained 
using sub-threshold optical parametric oscillators. 
Those are quite relevant schemes for quantum technology, 
since they allow one to generate entanglement and/or 
squeezed coherent light through amplitude and phase 
modulation of a signal encoded into spectral sideband 
modes \cite{barbosa:13,cialdi:16}.
\par
The paper is structured as follows. In Section~\ref{s:SNR} 
we review how to extract information about the moments of the
photon-number distribution of an optical Gaussian states from
the first two moments of suitable quadrature operators.
Though our interest is mostly on single-mode states, 
for the sake of completeness, we develop the theory in the
more general scenario of two-mode Gaussian states. The 
analytic results that we report in this section may be indeed
useful to investigate the performance of interferometric 
setups \cite{spara:15,RB:13,pede:14}. In Section~\ref{s:zero:g2}
we start with single-mode Gaussian states and seek a convenient
expression for their zero-delay time second-order correlation 
function. In particular, starting form the results of 
section~\ref{s:SNR}, we obtain an analytic expression of $g^{(2)} (0)$ 
as a function of the relevant parameters of the state, i.e. 
the coherent amplitude, the squeezing parameter and the mean number 
of thermal photons.  Then, we turn the attention to 
two-mode Gaussian states, and show that also the class of 
symmetric displaced squeezed thermal states
can exhibit a (two-mode)  $g^{(2)} (0) <1$. Remarkably, in both 
the two mentioned cases we find the same threshold value on the 
coherent amplitude leading to $g^{(2)} (0) <1$. The threshold value
depends on the other involved parameters, i.e. the single- or two-mode
complex squeezing parameter and the thermal contribution.
In section~\ref{s:exp} we describe the apparatus used to verify 
our theoretical predictions for the single-mode Gaussian states
and report the experimental results based on the analysis of the
homodyne traces. Section~\ref{s:concl} closes the paper with some
concluding remarks, also about future developments
concerning two-mode Gaussian states.
\section{Moments of the photon-number distribution}\label{s:SNR}
A Gaussian state is fully characterised by its covariance 
matrix (CM) $\bmsigma$ and first moments vector $\bmX$. 
For a single-mode Gaussian state $\varrho$ the
$2\times 2$ CM elements are given by 
\begin{subequations}
\begin{align}
[\bmsigma]_{qq} & = \langle \hq^2 \rangle - \langle \hq \rangle^2\,,\quad
[\bmsigma]_{pp}  = \langle \hp^2 \rangle - \langle \hp \rangle^2\,,\\
[\bmsigma]_{qp} & = \frac12\langle \hq \hp  +  \hp \hq \rangle -
\langle \hq \rangle  \langle \hp \rangle\,,
\end{align}
\end{subequations}
and first moments vector $\bmX = ( \langle \hq \rangle, \langle \hp \rangle)$,
where we introduced
the canonically conjugate operators $q = \hx_0$ and $p 
=\hx_{\pi/2}$, $\hx_\theta = (\ha^\dag\, e^{i\theta} + 
\ha\,e^{-i\theta} )/\sqrt{2}$ being the field-quadrature operator.
All the elements appearing in
$\bmsigma$ and $\bmX$ can be obtained by measuring suitable
 quadrature, e.g. by means
of a single homodyne detector. In particular, whereas it is 
clear how to obtain $\langle \hq \rangle$, $\langle \hp \rangle$,
$\sigma_{qq}$ and $\sigma_{pp}$ from the measurement of single quadrature moments, we note that
$\langle \hq \hp  + \hp \hq \rangle = \langle \hx_{\pi/4}^2 \rangle -\langle \hx_{-\pi/4}^2 \rangle$.
\par
Similar results can be obtained via a single homodyne 
detector for two-mode
Gaussian states \cite{buono:10, dauria:09} also for 
sideband modes \cite{barbosa:13, cialdi:16}.
In this case, $\bmsigma$ is the $4\times 4$ matrix CM, with
\begin{equation}
[\bmsigma]_{hk} = \frac12\langle \hat{R}_h\hat{R}_k + \hat{R}_k\hat{R}_h\rangle -
\langle\hat{R}_h\rangle\langle\hat{R}_k\rangle\,,
\end{equation}
and $\bmX = {\rm   Tr}[\varrho \hat{\bmR}]$, where $\hat{\bmR} = (\hq_1,\hp_1,\hq_2,\hp_2)$, with $\hq_k = (\ha_k^\dag + \ha_k)/\sqrt{2}$, $\hp_k = i(\ha_k^\dag - \ha_k)/\sqrt{2}$, and $\ha_k$ is the annihilation operator of the $k$-th mode, $[\ha_h,\ha_k^{\dag}]=\delta_{hk}$. The single-mode case is a particular case of the two-mode one, therefore in the following we develop the theory by focusing on the two-mode scenario. Furthermore, considering the two-mode case will allow us to obtain more general formula for intensity correlations useful for high precision, quantum-enhanced sensing \cite{RB:15, spara:15, RB:13, lopa:13}.
\par
Usually, the characterisation of optical quantum states or the measurement of tiny phase
fluctuations in interferometers may require to calculate the expectation values
of the products of the powers of the number operators, namely: $(\hN_1)^n (\hN_2)^m$,
where $\hN = \ha_k^{\dag}\ha_k$. These quantities can be expressed as
linear combinations of the symmetrically ordered products $[(\ha_k^\dag)^n \ha_k^m]_{\rm s}$
as follows (we stop at the 4-th order that is relevant in the most of practical cases
\cite{RB:13, lopa:13}):
\begin{subequations}\label{Nk:pow}
\begin{align}
\hN_k &= [\ha_k^\dag \ha_k]_{\rm s} - \frac12 \,, \\
(\hN_k)^2
&= [(\ha_k^\dag)^2 \ha_k^2]_{\rm s}
- [\ha_k^\dag \ha_k]_{\rm s} \,,\\
(\hN_k)^3
&=  [(\ha_k^\dag)^3 \ha_k^3]_{\rm s} -
\frac12 \left\{
3 [(\ha_k^\dag)^2 \ha_k^2]_{\rm s}
+ [\ha_k^\dag \ha_k]_{\rm s} - \frac12 \right\} \,,\\
(\hN_k)^4
&=  [(\ha_k^\dag)^4 \ha_k^4]_{\rm s}\notag\\
&\qquad- 2 \left\{[(\ha_k^\dag)^3 \ha_k^3]_{\rm s}
+  [(\ha_k^\dag)^2 \ha_k^2]_{\rm s}
+  [\ha_k^\dag \ha_k]_{\rm s}\right\} \,,
\end{align}
\end{subequations}
where $[(\ha_k^\dag)^n \ha_k^m]_{\rm s}$ may be obtained as \cite{cah:69}:
\begin{equation}
[(\ha_k^\dag)^n \ha_k^m]_{\rm s} =
\left.\frac{\partial^n_x \partial^{m}_y (x\, \ha_k^{\dag} + y\,
    \ha_k)^{n+m} }{(n+m)!}  \right|_{x=y=0}.
\end{equation}
Starting from the Eqs.~(\ref{Nk:pow}), we can write the quantity $(\hN_1)^n (\hN_2)^m$
as a linear combination of the products $[(\ha_1^\dag)^h \ha_1^h]_{\rm s} [(\ha_2^\dag)^k \ha_2^k]_{\rm s}$.
Therefore, the problem is now to connect the expectations of the symmetrically ordered
products to the CM $\bmsigma$ and the first moments vector $\bmX$. This can be
achieved by using the characteristic function of the state $\varrho$, namely,
\begin{equation}
\chi(\lambda_1,\lambda_2) = \Tr[D(\lambda_1) D(\lambda_2)\varrho]\,,
\end{equation}
with $D(\lambda_k) = \exp(\lambda_k\ha_k^{\dag} - \lambda_k^* \ha_k)$,
$\lambda_k \in {\mathbbm C}$,
which can be recast in the following Gaussian form:
\begin{equation}\label{chi:in}
\chi(\bmLambda) = \exp\left\{
-\mbox{$\frac12$} \bmLambda^{T}\bmsigma\,\bmLambda + i
\bmLambda^{T}\bmX
\right\}\,,
\end{equation}
where $\bmLambda^{T}=(\sfx_1,\sfy_1,\sfx_2,\sfy_2)$.
Now we can use $\chi(\bmLambda)$ as the moment-generating function to calculate
the moments by using the following identity \cite{FOP:05}:
\begin{equation}\label{moments:s}
\langle [(\ha_1^\dag)^h \ha_1^h]_{\rm s}
[(\ha_2^\dag)^k \ha_2^k]_{\rm s} \rangle =
\partial_{\lambda_1}^k \partial_{\lambda_1^*}^k
\partial_{\lambda_2}^h \partial_{\lambda_2^*}^h\,
\chi(\lambda_1,\lambda_2)\Big|_{\lambda_1=\lambda_2=0}\,.
\end{equation}
For the sake of clarity, it is useful to rewrite the $4\times 4$ CM $\bmsigma$ and the mean values vector $\bmX$ as follows
(we highlighted the upper-left and lower-right $2 \times 2$ blocks which refer to the CM matrices of the reduced single-mode states
$\varrho_h = \Tr_k[\varrho]$, $h\ne k$, of mode $\ha_h$, $h,k=1,2$):
\begin{equation}
\bmsigma = 
\left(
\begin{array}{cc|cc}
a & c & e & f \\
c & b & g & h \\
\hline
e & g & A & C \\
f & h & C & B
\end{array}
\right)\,,\quad
\bmX^T = (X_1,Y_1,X_2,Y_2).
\end{equation}
Therefore, the characteristic function in Eq.~(\ref{chi:in}) can be expressed in the complex notation as \cite{oli:07}:
\begin{align}
\chi(\lambda_1,\lambda_2)
=\,& \exp\Big\{
-{\mathscr A}|\lambda_1|^2 -
{\mathscr B}|\lambda_2|^2  \nonumber \\[1ex]
&- {\mathscr C}\lambda_1^2 -
{\mathscr C}^*{\lambda_1^*}^2 -
{\mathscr D}\lambda_2^2 -
{\mathscr D}^*{\lambda_2^*}^2 \nonumber \\[1ex]
&-{\mathscr E}\lambda_1\lambda_2 -
{\mathscr E}^*\lambda_1^* \lambda_2^* -
{\mathscr F}\lambda_1 \lambda_2^* -
{\mathscr F}^*\lambda_1^* \lambda_2 \nonumber \\[1ex]
&+ i\big[{\mathscr U}^*\lambda_1 + {\mathscr U}\lambda_1^* +
{\mathscr V}^*\lambda_2 + {\mathscr V}\lambda_2^*\big]\Big\},
\label{chi:comp}
\end{align}
where:
\begin{subequations}
\begin{align}
\label{cal:mode:1}
&{\mathscr A} = \frac{a+b}{2},\quad
{\mathscr C} = \frac{a-b-2ic}{4},\quad
{\mathscr U} = \frac{X_1 + i Y_1}{\sqrt{2}},\\[1ex]
&{\mathscr B} = \frac{A+B}{2},\quad
{\mathscr D} = \frac{A-B-2iC}{4},\quad
{\mathscr V} = \frac{X_2 + i Y_2}{\sqrt{2}},\\[1ex]
&{\mathscr E} = \frac{e-h-i(f+g)}{2},\quad
{\mathscr F} = \frac{e+h+i(f-g)}{2}.
\end{align}
\end{subequations}
By using Eqs.~(\ref{chi:comp}) and (\ref{moments:s})
we can obtain the expressions for the symmetrically ordered moments
up to the 2-nd order which we will use in the next section (the analytic formulas
for higher orders are cumbersome and are not reported here):
\begin{subequations}
\label{symm:moments:1}
\begin{align}
&\langle [a_1^\dag a_1]_{\rm s} \rangle = {\mathscr A}+|{\mathscr U}|^2\,,\\[1ex]
&\langle [(a_1^\dag)^2 a_1^2]_{\rm s} \rangle =
2{\mathscr A}^2 + 4 {\mathscr A}|{\mathscr U}|^2
+|{2\mathscr C}+{\mathscr U}^2|^2\,,
\end{align}
\end{subequations}
which refer to mode 1 while, analogously, for the mode 2 we have:
\begin{subequations}
\begin{align}
&\langle [a_2^\dag a_2]_{\rm s} \rangle = {\mathscr B}+|{\mathscr V}|^2\,,\\[1ex]
&\langle [(a_2^\dag)^2 a_2^2]_{\rm s} \rangle =
2{\mathscr B}^2 + 4 {\mathscr B}|{\mathscr V}|^2
+|{2\mathscr D}+{\mathscr V}^2|^2\,,
\end{align}
\end{subequations}
and
\begin{align}
\langle [a_1^\dag a_1]_{\rm s} [a_2^\dag a_2]_{\rm s} \rangle = \, &
|{\mathscr E}|^2 + |{\mathscr F}|^2 + |{\mathscr U}|^2\,|{\mathscr V}|^2 \nonumber\\
&+ {\mathscr A}|{\mathscr V}|^2 +
{\mathscr B}|{\mathscr U}|^2 + {\mathscr A}{\mathscr B} \nonumber\\
&+ {\mathscr U}^*{\mathscr V}^*{\mathscr E} +
{\mathscr U}{\mathscr V}{\mathscr E}^* \nonumber\\
&+  {\mathscr V}^*{\mathscr U}{\mathscr F} +
{\mathscr F}^*{\mathscr U}^*{\mathscr V}\,.
\end{align}
that is connected to the correlations between the modes.

\section{Second-order correlation function}\label{s:zero:g2}

\subsection{Single-mode Gaussian states}
In this section we apply the analytic results obtained above to study the zero-delay time
second-order correlation function of a single-mode Gaussian state, namely:
\begin{subequations}
\begin{align}
g^{(2)}(0) &= \frac{\langle \ha^{\dag} \ha^{\dag} \ha \ha \rangle}{\langle \ha^{\dag} \ha \rangle^2}\,
\\[1ex]
&= \frac{\langle (\ha^{\dag} \ha)^2 \rangle - \langle \ha^{\dag} \ha \rangle}{\langle \ha^{\dag} \ha \rangle^2}\,
\\[1ex]
&= \,\frac{2\left( 2\langle [(\ha^\dag)^2 \ha^2]_{\rm s} \rangle - 4 \langle [(\ha^\dag) \ha]_{\rm s} \rangle  + 1 \right)}
{\left( 2 \langle [(\ha^\dag) \ha]_{\rm s} \rangle  - 1 \right)^2}\,, \label{g:2:CM:elem}
\end{align}
\end{subequations}
where $\ha$ is the considered boson field mode. It is thus clear that Eq.~(\ref{g:2:CM:elem})
can be evaluated just  starting from the CM elements and mean values of the (Gaussian) input
state.
\par
The density operator of the most general single-mode Gaussian state (sGs) can be always written
as \cite{adam:95,FOP:05}:
\begin{equation}
\varrho= D(\alpha)S(\xi)\,\nu(N_{\rm th})\,S^{\dag}(\xi) D^{\dag}(\alpha) \,,
\label{gen:sm:GS}
\end{equation}
where $D(\alpha)$ is the displacement operator introduced above, $S(\xi) = \exp[\frac12\xi (\ha^{\dag})^2 - \frac12 \xi^* \ha^2]$
is the squeezing operator and $\nu(N_{\rm th}) = (N_{\rm th})^{\ha^\dag \ha} / (1+ N_{\rm th})^{\ha^\dag \ha + 1}$ is
a ``thermal'' state with $N_{\rm th}$ mean photons. The explicit expressions of the CM elements associated with $\varrho$
as functions of $N_{\rm th}$ and $\xi = r\, e^{i\psi}$ are:
\begin{subequations}
\begin{align}
\sigma_{qq}&= \frac{1+2N_{\rm th}}{2}\:
\Big[\!\cosh(2r)-\sinh(2r)\cos\psi\Big] \:, \\[1ex]
\sigma_{pp}&= \frac{1+2N_{\rm th}}{2}\:
\Big[\!\cosh(2r)+\sinh(2r)\cos\psi\Big] \:, \\[1ex]
\sigma_{qp}&=\sigma_{21}=\frac{1 + 2N_{\rm th}}{2}\:
\sinh(2r)\sin\psi \:,
\end{align}
\end{subequations}
while the first-moments vector is $\bmX = \sqrt{2}(\Re\mbox{e}[\alpha],\Im\mbox{m}[\alpha])$.
Without loss of generality we can assume $\alpha \in {\mathbbm R}$.
By using Eqs.~(\ref{cal:mode:1}) and (\ref{symm:moments:1})
for the mode $\hat{a}$, Eq.~(\ref{g:2:CM:elem}) leads to the following expression
for the second order correlation function:
\begin{align}
g^{(2)}_{\rm sGs}(0) = \, & 2 + \Big\{
(1+ 2 N_{\rm th}) \sinh(2r)  \nonumber \\
& \times\Big[(1+ 2 N_{\rm th}) \sinh(2r) +4 \alpha^2 \cos\psi\Big]- 4 \alpha^4 \Big\} \nonumber\\
& \times\Big[  (1+ 2 N_{\rm th}) \cosh (2r) +2 \alpha^2 -1 \Big]^{-2} \,.
\end{align}
It is worth noting that $g^{(2)}_{\rm sGs}(0)$ can assume values lass than 1 and, in particular, we focus on
the dependence of $g^{(2)}_{\rm sGs}(0)$ on the displacement amplitude $\alpha$, in fact, the state
(\ref{gen:sm:GS}) can be always seen as the result of the interference of between a squeezed vacuum
and a coherent state at a beam splitter with suitable transmissivity (where only a single output port is
considered).
\par
Straightforward calculations show that there is a threshold value $\alpha_{\rm Th}(r,\psi,N_{\rm th})$
such that $g^{(2)}_{\rm sGs}(0)=1$, and it reads:
\begin{align}
&\alpha_{\rm Th}(r,\psi,N_{\rm th}) = \nonumber\\
& \frac12\sqrt{\frac{(1+ 2 N_{\rm th})^2 \sinh^2(2r) +\left[(1+ 2 N_{\rm th}) \cosh(2r) -1\right]^2}
{{\cal T}(r,N_{\rm th}) - (1+ 2 N_{\rm th}) \sinh(2r) (1+\cos\psi)}}\,,\label{a:th:ncD}
\end{align}
where we introduced ${\cal T}(r,N_{\rm th}) = 1 - (1+ 2 N_{\rm th})
e^{-2r}$: if ${\cal T}(r,N_{\rm th}) > 0$, then ${\cal T}(r,N_{\rm th})$
corresponds to the nonclassical depth of the state (\ref{gen:sm:GS})
\cite{nonc:depth,brunelli:15}, namely, a measure of the nonclassicality
of the state.  Since we assumed $\alpha$ to be real, the previous
threshold exist iff its denominator is positive.
\begin{figure}[h!]
\begin{center}
\includegraphics[width=0.8\columnwidth]{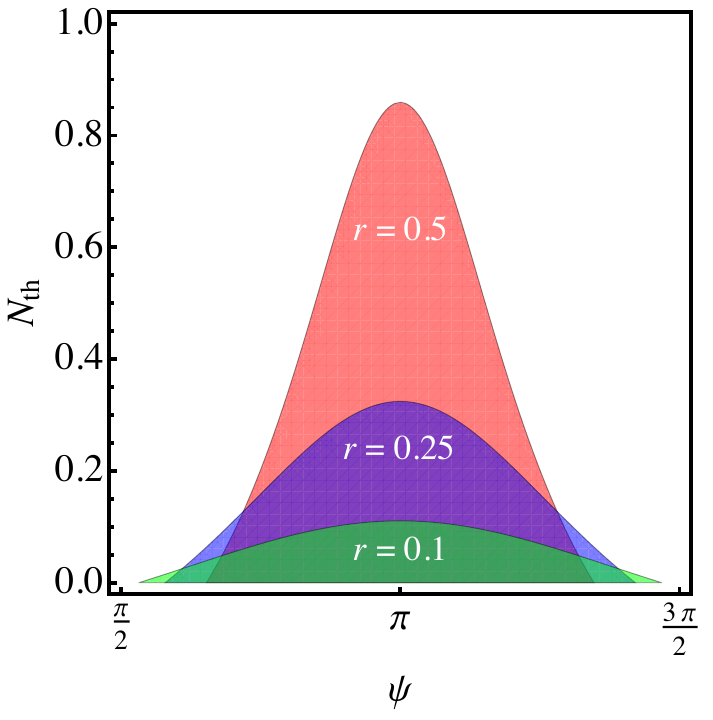}
\vspace{-0.8cm}
\end{center}
\caption{The threshold condition on the coherent amplitude of a single-mode
Gaussian state to have $g^{(2)}_{\rm sGs}(0) < 1$. The shaded areas (each 
one corresponding  to a different value of the squeezing parameter $r$)
are the regions of the $\psi$--$N_{\rm th}$ plane where a threshold 
$\alpha_{\rm Th}(r,\psi,N_{\rm th})$ 
exists. For values of $\psi$ and $N_{\rm th}$ outside those regions
we have $g^{(2)}_{\rm sGs}(0) >1$.}\label{f:condition}
\end{figure}
\par
In Fig.~\ref{f:condition}, we show the regions of the $\psi$--$N_{\rm th}$
plane where this positivity condition is satisfied. When $\alpha_{\rm
Th}(r,\psi,N_{\rm th})$
exists, we have (for a given set of parameters $\{r,\psi,N_{\rm th}$)
that $g^{(2)}_{\rm sGs}(0) < 1$ if $\alpha > \alpha_{\rm
Th}(r,\psi,N_{\rm th})$ and {\it vice versa}. If, on the other hand, the
threshold does not exists, then the second-order correlation function is
always greater than 1. The threshold, for a given $r$, may exist only if
$\pi/2 \le \psi \le 3\pi/2$ and, as the value of $r$ increases, the
shaded region becomes more and more peaked around $\psi = \pi$ (see
Fig.~\ref{f:condition}).
\par
It is worth noting that for $\psi = \pi$ (amplitude squeezing), Eq.~(\ref{a:th:ncD}) reduces to:
\begin{align}
&\alpha_{\rm Th}(r,\pi,N_{\rm th}) = \nonumber\\
& \frac12\sqrt{\frac{(1+ 2 N_{\rm th})^2 \sinh^2(2r) +\left[(1+ 2 N_{\rm th}) \cosh(2r) -1\right]^2}
{{\cal T}(r,N_{\rm th})}}\,, \label{a:th:ncD:pi}
\end{align}
therefore, in the presence of amplitude squeezing and real displacement, {\it the
zero-delay second-order correlation function can be less the 1 iff the state
is nonclassical, namely, exhibits a non-null nonclassical depth ${\cal T}(r,N_{\rm th})$}. Otherwise,
$\alpha_{\rm Th}(r,\pi,N_{\rm th})$ does not exist (we recall that, according to
our choice of the parameters, this threshold should be real).
In the this case ($\psi = \pi$), the minimum (if exists!) occurs at (see also \cite{alex:16,pede:14}):
\begin{align}
\alpha_{\rm min}(r,\pi,N_{\rm th}) &= \sqrt{(1 + 2 N_{\rm th}) \sinh(2r)} \nonumber \\
&\hspace{1.0cm}\times \sqrt{\frac{(1 + 2 N_{\rm th}) \sinh(2r)}{{\cal T}(r,N_{\rm th})}-\frac12}\,.
\end{align}
\par
\begin{figure}[h!]
\begin{center}
\includegraphics[width=0.9\columnwidth]{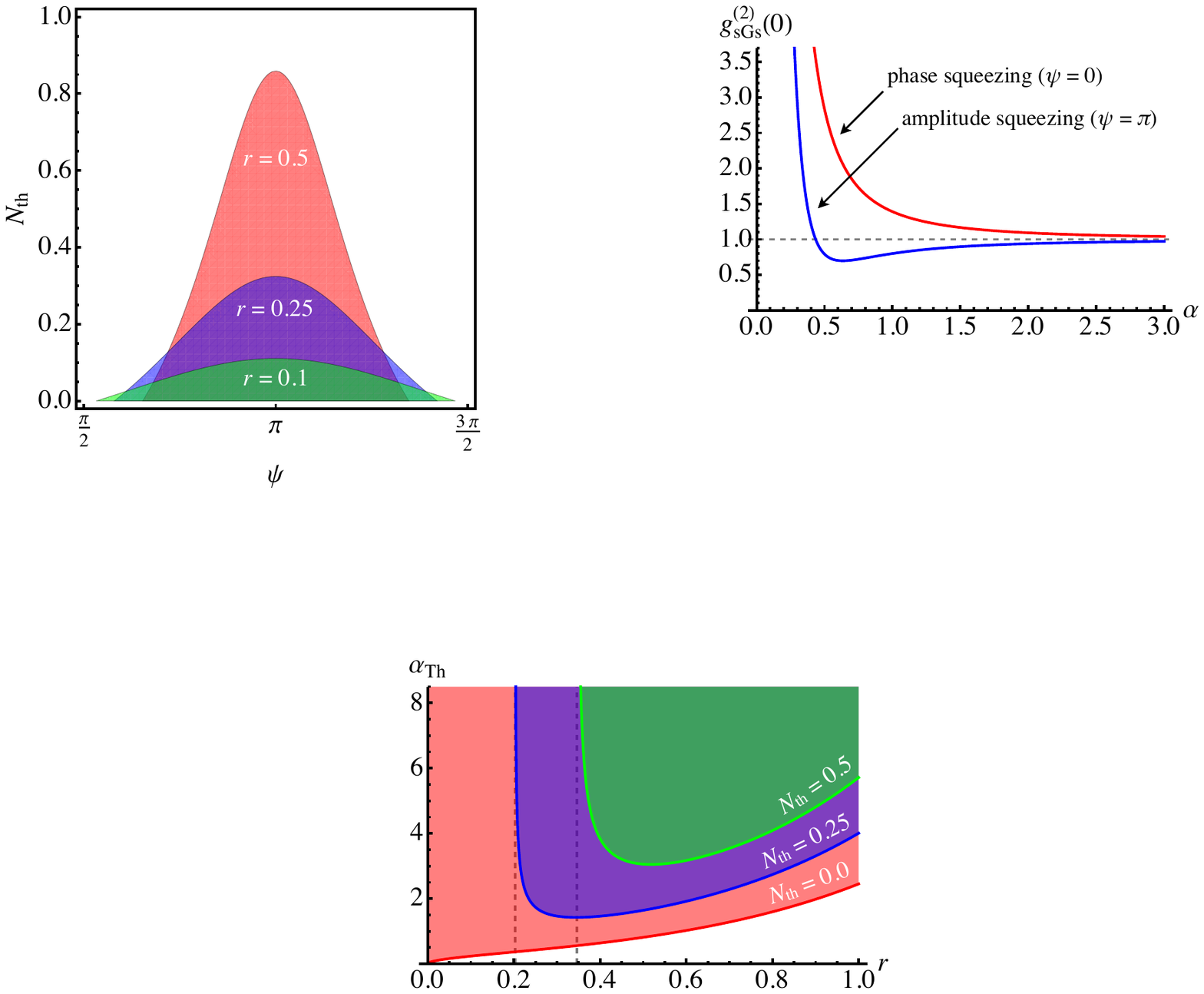}
\end{center}
\vspace{-0.5cm}
\caption{The threshold on the amplitude 
$\alpha_{\rm Th}(r,\pi,N_{\rm th})$ (for $\psi = \pi$) as 
a function of $r$ for different values of $N_{\rm th}$.
In the shaded regions $g^{(2)}_{\rm sGs}(0) <1$. The dashed vertical
lines correspond to $r_{\rm Th}( N_{\rm th}) = \frac12 \log(1 +  2
N_{\rm th})$: if $r \le r_{\rm Th}(N_{\rm th})$ the single-mode Gaussian
state is classical (it has zero nonclassical depth) and
$\alpha_{\rm Th}$ does not exist.}\label{f:ath:r}
\end{figure}
In Fig.~\ref{f:ath:r} we plot the threshold $\alpha_{\rm Th}(r,\pi,N_{\rm th})$ and highlight
the regions in which $g^{(2)}_{\rm sGs}(0) <1$: we can see that as the value of $N_{\rm th}$
increases, thus reducing the purity $\mu[\varrho] = (1 + 2 N_{\rm th})^{-1}$ of the state
(\ref{gen:sm:GS}), $\alpha_{\rm Th}(r,\pi,N_{\rm th})$ increases as well.

\subsection{Two-mode Gaussian states}
Now we consider the second-order correlation function of two-mode Gaussian states,
which, using Eqs.~(\ref{Nk:pow}), can be written as:
\begin{subequations}
\begin{align}
g^{(2)}_{\rm TM}(0) =& \frac{\langle \ha^{\dag} \ha^{\dag} \ha \ha \rangle +
\langle \hb^{\dag} \hb^{\dag} \hb \hb \rangle + 2 \langle \ha^{\dag} \ha\, \hb^{\dag} \hb \rangle}
{\left( \langle \ha^{\dag} \ha \rangle +  \langle \hb^{\dag} \hb \rangle \right)^2}\,
\\[1ex]
=& \,\frac{ \langle [(\ha^\dag)^2 \ha^2]_{\rm s} \rangle + \langle [(\hb^\dag)^2 \hb^2]_{\rm s} \rangle}
{\left[ \langle [(\ha^\dag) \ha]_{\rm s} \rangle + \langle [(\hb^\dag) \hb]_{\rm s} \rangle  - 1 \right]^2} \nonumber\\[1ex]
&\hspace{0.5cm} - \frac{3\left( \langle [\ha^\dag \ha]_{\rm s} \rangle + \langle [\hb^\dag \hb]_{\rm s} \rangle  -\frac12 \right)}
{\left[ \langle [(\ha^\dag) \ha]_{\rm s} \rangle + \langle [(\hb^\dag) \hb]_{\rm s} \rangle  - 1 \right]^2} \nonumber\\[1ex]
&\hspace{0.5cm} + \frac{2 \langle [\ha^\dag \ha]_{\rm s} [\hb^\dag \hb]_{\rm s} \rangle}
{\left[ \langle [(\ha^\dag) \ha]_{\rm s} \rangle + \langle [(\hb^\dag) \hb]_{\rm s} \rangle  - 1 \right]^2}\,,
\label{g:2:CM:elem:tm}
\end{align}
\end{subequations}
where $\ha = \ha_1$ and $\hb = \ha_2$ are the two involved modes.
Here, for the sake of clarity, we restrict or analysis to the class of the two-mode squeezed
thermal states (TMSTS) which we can generate and manipulate with the current technology
\cite{dauria:09,cialdi:16}. The density operator associated with a TMSTS can be written as:
\begin{equation}\label{TMSTS}
\varrho_{ab} =
D_{ab}(\alpha,\beta)\,S_2(\xi)\
\nu_a(N_{\rm th, 1})\otimes \nu_b(N_{\rm th, 2})\,S_2^{\dag}(\xi)
D_{ab}^{\dag}(\alpha,\beta) \,,
\end{equation}
where $D_{ab}(\alpha,\beta)=D_a(\alpha) \otimes D_b(\beta)$, with $D_k(z) = \exp(z \hk^\dag - z^* \hk)$,
$S_2(\xi) = \exp(\xi \ha^\dag \hb^\dag - \xi^* \ha \hb)$ is the two-mode squeezing operator
($\xi = r\, e^{i \psi}$ is now the two-mode squeezing parameter) and
$\nu_k(N) = N^{\hk^\dag \hk} / (1+ N)^{\hk^\dag \hk + 1}$ is the ``thermal'' state
of mode $\hk$ with $N$ mean photons. The corresponding $4\times 4$ CM has the following
block matrix form \cite{FOP:05}:
\begin{equation}\label{sigma:TM}
\bmsigma = \frac12 \left(\begin{array}{cc}
A\, \iid_2 & C\, \bmR_{\xi}\\[1ex]
C\,\bmR_{\xi} & B\, \iid_2
\end{array}\right)
\end{equation}
where $\iid$ is the $2\times 2$ identity matrix,
\begin{equation}
\bmR_{\xi} = \sinh r\, \left(\begin{array}{cc}
\cos\psi & \sin\psi\\[1ex]
\sin\psi & -\cos\psi
\end{array}\right)
\end{equation}
and:
\begin{subequations}
\begin{align}
A &= \cosh (2r) + 2\left( N_{\rm th, 1} \cosh^2 r + N_{\rm th, 2} \sinh^2 r \right)\,, \\[1ex]
B &= \cosh (2r) + 2\left( N_{\rm th, 1} \sinh^2 r + N_{\rm th, 2} \cosh^2 r \right)\,, \\[1ex]
C &= (1 + N_{\rm th, 1} + N_{\rm th, 2}) \sinh (2r)\,.
\end{align}
\end{subequations}
\par
Given the CM (\ref{sigma:TM}) it is straightforward to calculate the second-order correlation
function $g^{(2)}_{\rm TM}(0)$ of the corresponding state $\varrho_{ab}$ in Eq.~(\ref{TMSTS}).
The analytical formula is clumsy and it is not reported explicitly in its general form.
However, the results are similar to those obtained in the case of the single-mode Gaussian states
discussed above. In particular, if we set $\psi = \pi$ we can still find the threshold values of the coherent amplitudes
$\alpha$ and $\beta$ and of the thermal contributions $N_{\rm th, 1}$ and $N_{\rm th, 2}$, in order to have
$g^{(2)}_{\rm TM} < 1$. This is not possible if we choose $\psi = 0$ instead.

\par
\begin{figure}[h!]
\begin{center}
\includegraphics[width=0.98\columnwidth]{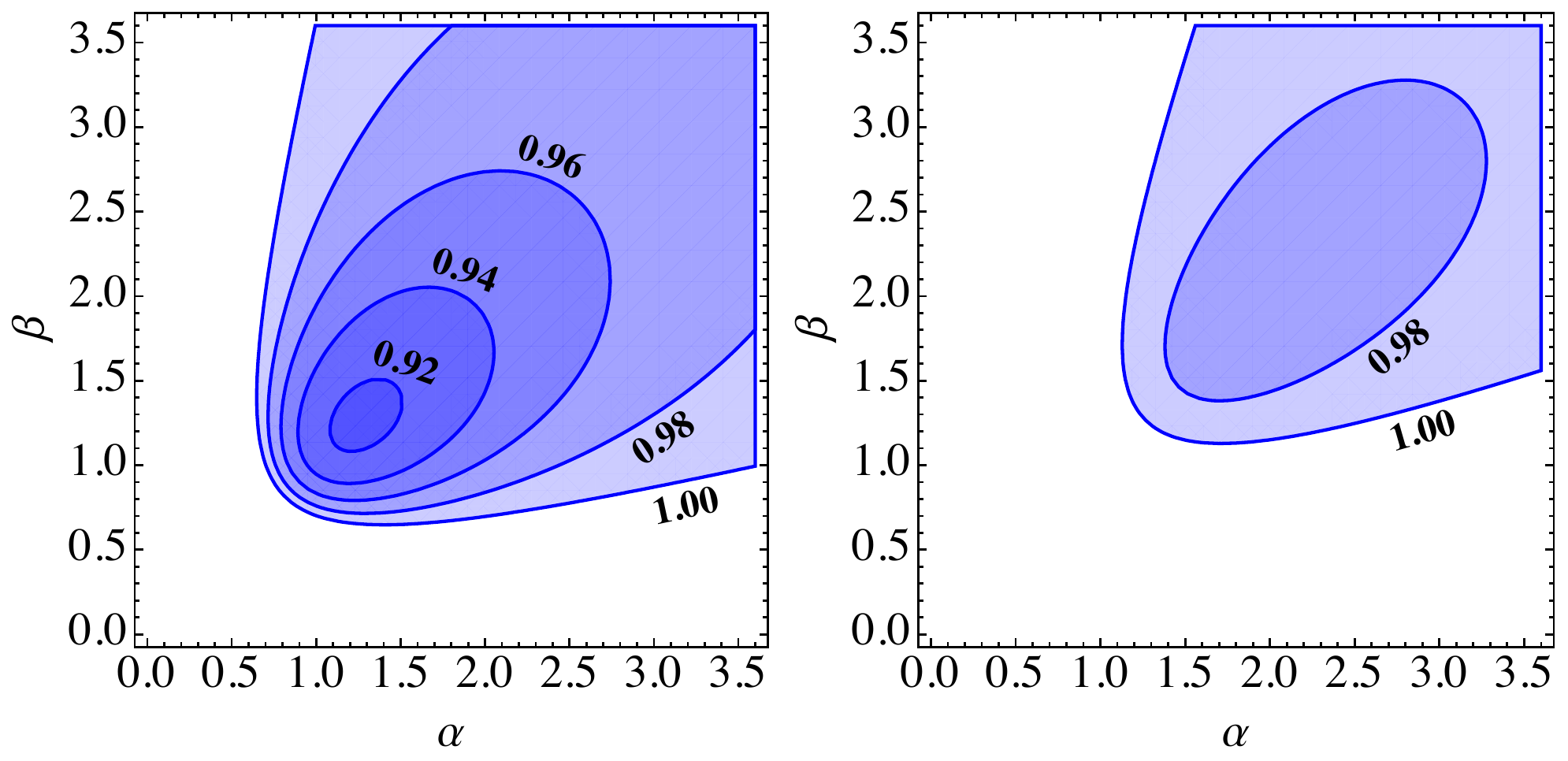}
\end{center}
\vspace{-0.5cm}
\caption{
Contour plot of the the second-order correlation function $g^{(2)}_{\rm TM}$ of two-mode Gaussian states
as a function of the displacement amplitudes $\alpha$ and $\beta$ in the case of the TMSTS
given in Eq.~(\ref{TMSTS}) with $N_{\rm th,1} = N_{\rm th,2} = 0$ (left panel)
and $N_{\rm th,1} = N_{\rm th,2} = 0.15$ (right panel). In both the panels we set $r=0.5$ and $\psi = \pi$, where $\xi = r\, e^{i\psi}$
is the two-mode squeezing parameter, see the text for details.
Only the region for $g^{(2)}_{\rm TM} < 1$ is shown
(note that for $\psi = 0$ one has $g^{(2)}_{\rm TM} > 1$, $\forall \alpha, \beta \ge 0$).}
\label{f:tm:dis}
\end{figure}
\begin{figure}[h!]
\begin{center}
\includegraphics[width=0.98\columnwidth]{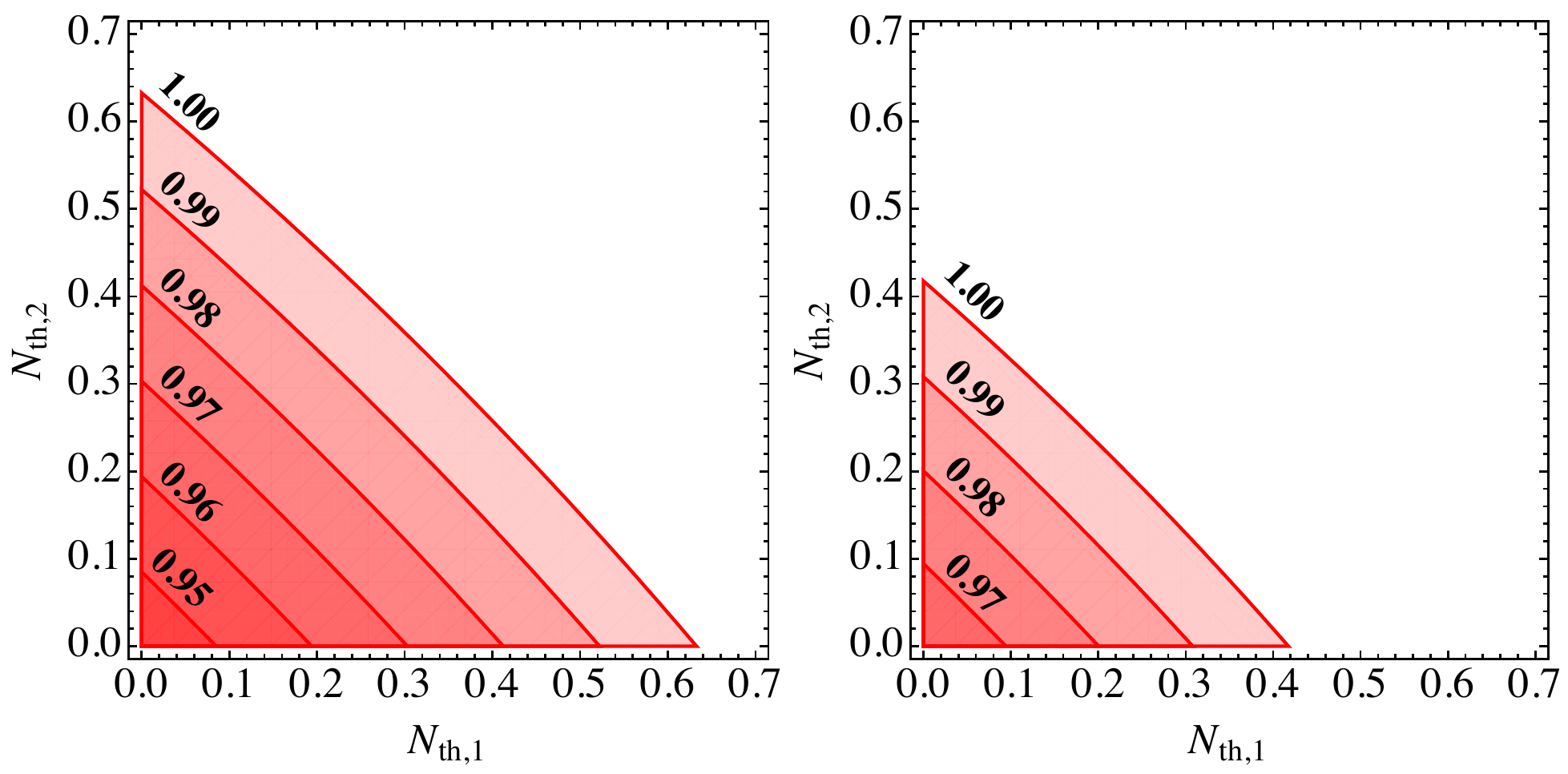}
\end{center}
\vspace{-0.5cm}
\caption{
Contour plot of the the second-order correlation function $g^{(2)}_{\rm TM}$ of two-mode Gaussian states
as a function of the thermal contributions $N_{\rm th,1}$ and $N_{\rm th,2}$ in the case of the TMSTS
given in Eq.~(\ref{TMSTS}) with $\alpha = \beta = 2$  and two-mode squeezing parameter $r=0.5$ (left panel)
and $r=0.2$ (right panel), where $\xi = r\, e^{i\psi}$ and we set $\psi = \pi$ in both the panels, see the text for details.
Only the region for $g^{(2)}_{\rm TM} < 1$ is shown
(note that for $\psi = 0$ one has $g^{(2)}_{\rm TM} > 1$, $\forall N_{\rm th,1}, N_{\rm th,2} \ge 0$).
}\label{f:tm:nth}
\end{figure}

In Fig.~\ref{f:tm:dis} we report the region of the plane $(\alpha, \beta)$ leading to $g^{(2)}_{\rm TM} < 1$
for two values of the thermal contribution (for the sake of simplicity we set $N_{\rm th,1} = N_{\rm th,2}$)
and $r=0.5$ (with $\psi = \pi$, as mentioned above) that is a typical value we can easily reach in the experiments.
Figure~\ref{f:tm:nth} shows the region of the plane $(N_{\rm th,1}, N_{\rm th,2})$ for which $g^{(2)}_{\rm TM} < 1$
when we fix $\alpha=\beta=2$ and use two values of the two-mode squeezing parameter (its phase $\psi$ is
still set equal to $\pi$).
\par
From the Figs.~\ref{f:tm:dis} and \ref{f:tm:nth} its is clear that the best working regime is obtained in the presence of
a symmetric configuration, namely, for $\alpha = \beta$ and $N_{\rm th,1} = N_{\rm th,2}$. This is usually the regime
achieved in actual experiments involving sideband modes \cite{cialdi:16}. In this case, it is possible to write
a more compact analytical expression for the two-mode second-order correlation function and we can find
the following threshold of $\alpha$  such that $g^{(2)}_{\rm TM}(0)=1$, namely:
\begin{align}
&\tilde{\alpha}_{\rm Th}(r,\psi,N_{\rm th}) = \nonumber\\
& \frac12\sqrt{\frac{(1+ 2 N_{\rm th})^2 \sinh^2(2r) +\left[(1+ 2 N_{\rm th}) \cosh(2r) -1\right]^2}
{\frac12 \left[ {\cal T}(r,N_{\rm th}) + {\cal T}(-r,N_{\rm th}) \right] - (1+ 2 N_{\rm th}) \sinh(2r) \cos\psi}}\,,\label{a:th:ncD:TM}
\end{align}
where ${\cal T}(\pm r,N_{\rm th})$ was introduced in Eq.~(\ref{a:th:ncD}).
Note that for $\psi = \pi$, the threshold $\tilde{\alpha}_{\rm Th}(r,\pi,N_{\rm th})$ reduces to the same threshold
obtained for the single-mode Gaussian states given in Eq.~(\ref{a:th:ncD:pi}).

\par
Up to now, we have studied $g^{(2)}(0)$ as a function of the relevant parameters
characterising the single- or two-mode Gaussian state under investigation.
However, if we focus on the single-mode states, thanks to Eqs.~(\ref{cal:mode:1}) and
(\ref{symm:moments:1}) we can retrieve the value of $g^{(2)}_{\rm sGs}(0)$ by
acquiring the information about the covariance matrix and the
first moment vector. These quantities straightforwardly follow from the measurement
of the four quadratures $\hq$, $\hp$ and $\hx_{\pm \pi/4}$ as mentioned
in the section~\ref{s:SNR} and they can be experimentally obtained by a
single homodyne detector, as we are going to demonstrate in the next section.
Similar results can be obtained for two-mode Gaussian states \cite{buono:10,cialdi:16},
and they are worth to be thoroughly investigated in future works.

\section{Experimental apparatus and single-mode state results}\label{s:exp}
In order to test the theoretical previsions, we built an experimental 
setup to generate and manipulate displaced-squeezed states. In particular,
our scheme allows controlling both the coherent amplitude and the squeezing
parameter of the states as well as their relative phase. Therefore we can 
process the two families of amplitude and phase squeezed states.
\par
\begin{figure}[h!]
\includegraphics[width=0.99\columnwidth]{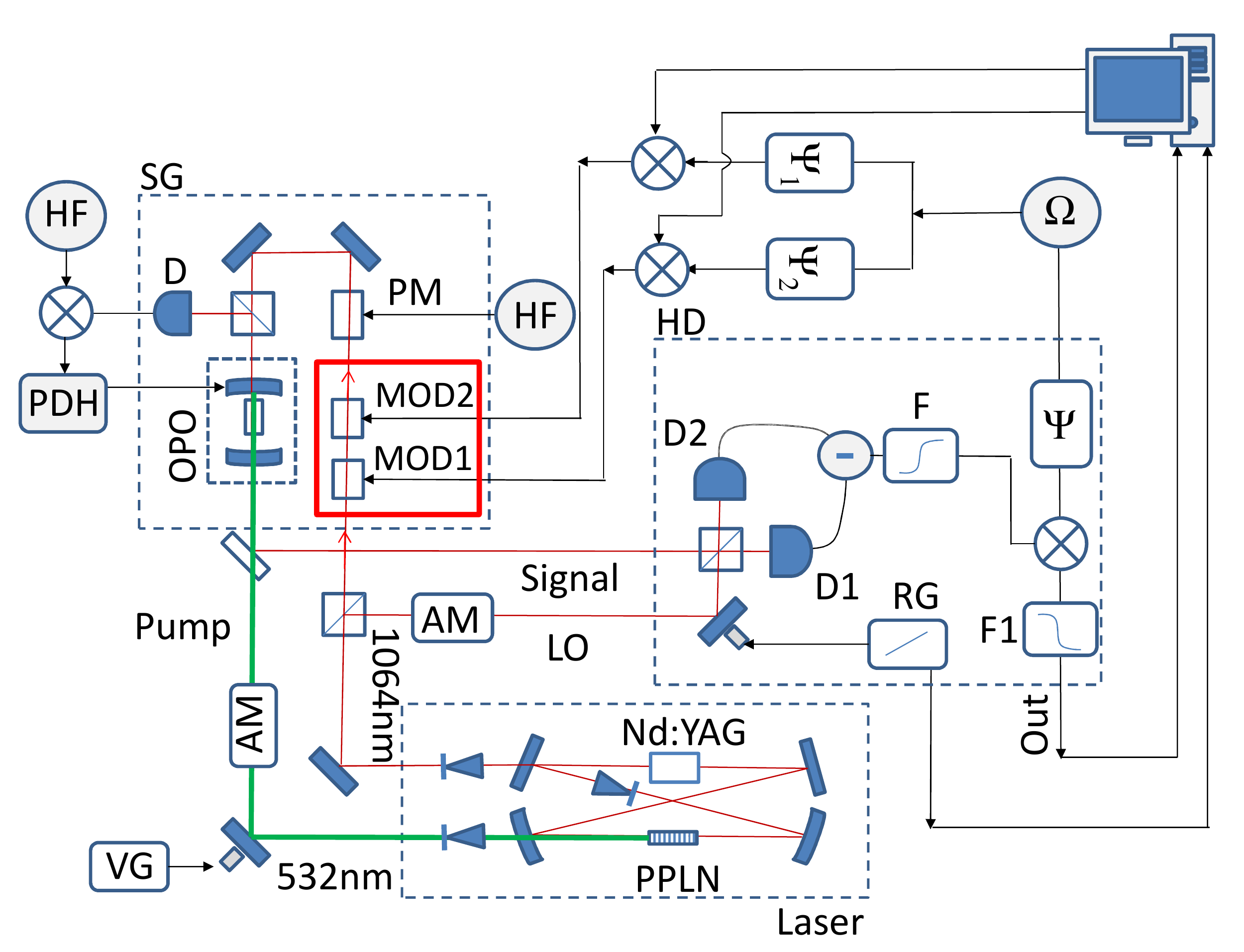}
\caption{(Color online) Schematic diagram of the experimental setup 
to generate squeezed states. 
The principal radiation source is provided by a home made Nd:YAG Laser
internally-frequency-doubled.  One laser output
(@~532~nm) pumps the MgO:LiNbO$_3$ crystal (length 10~mm) of the optical
parametric oscillator (OPO) whereas the other output (@~1064~nm) is sent
to a polarising beam splitter to generate the local oscillator
(LO) as well as the seed signal for the OPO. 
The power of the LO is set by an amplitude modulator (AM). The phase modulators (PM) 
generates  the sidebands @~110~MHz for the active stabilisation of the OPO cavity. 
Whereas the two modulators (MOD1, MOD2) generate the sidebands @~3~MHz
for the seed state generation, in particular by exploiting these two modulators we can
set the values of $\alpha$ and $\psi$ by the computer.
The length of the OPO cavity is actively controlled by a piezo connected to its rear mirror.
The homodyne detector consists of a 50:50 beam splitter,  two low noise detectors (D1, D2)
and a differential amplifier.
The relative phase between the pump @~532~nm and the radiation @~1064~nm
is set by a piezo connected with a voltage generator (VG).}
\label{f:schema}
\end{figure}
Figure~\ref{f:schema} shows the scheme of our experimental apparatus.
It consists of three stages: Laser, signal generator (SG) and homodyne detector (HD) (see more details in \cite{fidelity}).
In particular continuous-wave squeezed light is generated by a sub-threshold optical parametric oscillator (OPO). The OPO input seed (1064~nm) and the OPO pump beam (532~nm) arise from a home-made internally frequency doubled Nd:YAG laser. The output at 1064~nm is split into two beams by using a polarising beam splitter (PBS): one is used as the local oscillator (LO) for the homodyne detector and the other is sent into the OPO. A phase modulator (PM) generates a signal at frequency of 110~MHz (HF) used as active stabilisation of the OPO cavity via the Pound-Drever-Hall (PDH) technique \cite{cialdi:16}. In order to generate the coherent squeezed states our strategy is to exploit the combined effect of two optical modulators (MOD1, MOD2) placed before the OPO. By properly chosen the modulation amplitudes \cite{fidelity}, it is possible to generate arbitrary coherent states on the sidebands @~3~MHz for seeding the OPO \cite{cialdi:16,fidelity}. In this way we can set both the value of the phase $\psi$ and the 
value of $\alpha$ on demand by the computer.
\begin{figure}[h!]
\includegraphics[width=0.99\columnwidth]{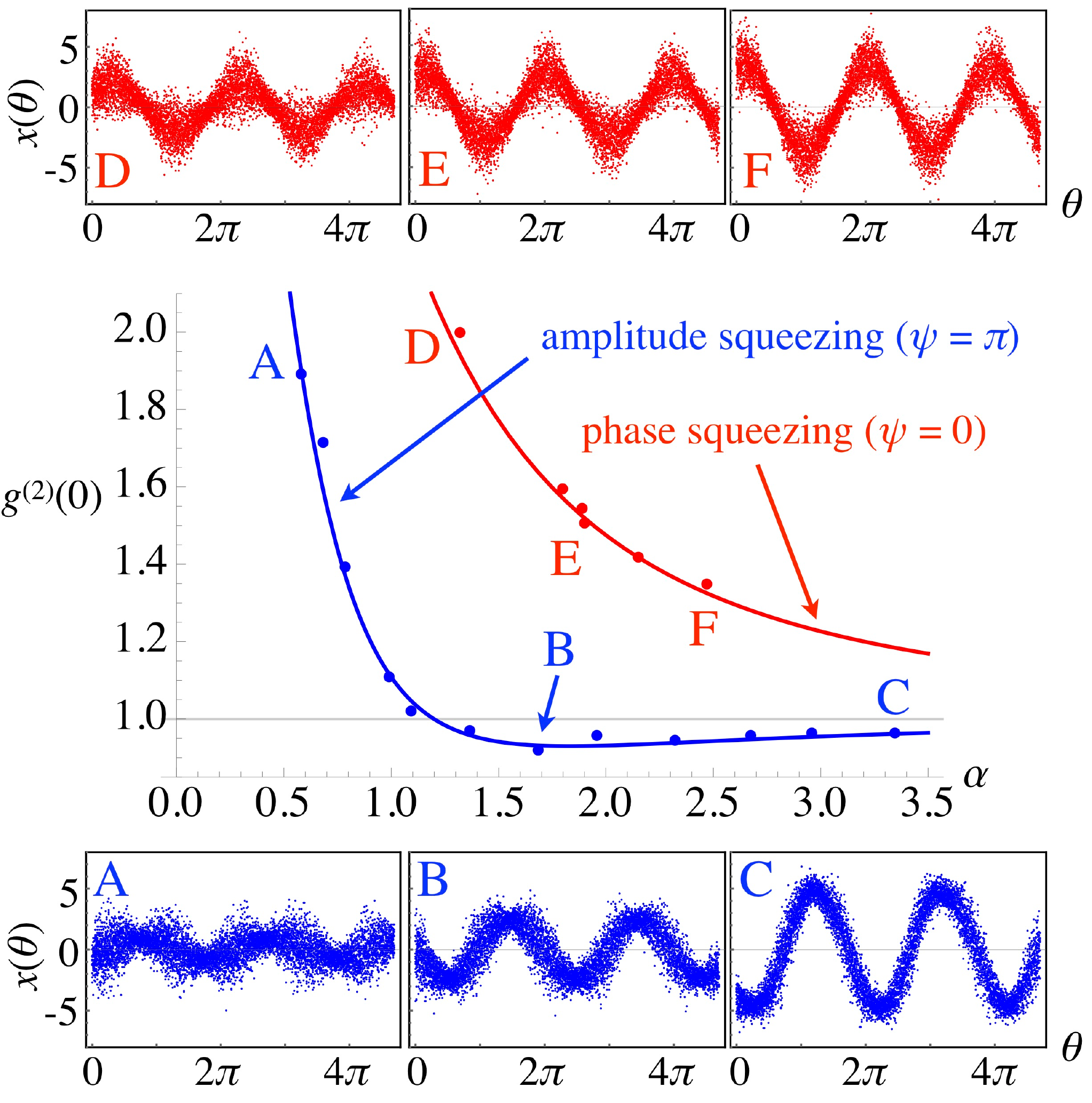}
\caption{Experimental $g^{(2)}_{\rm sGs}(0)$ (dots) obtained through Eq.~(\ref{g:2:CM:elem})
in the case of amplitude ($r = 0.5 \pm 0.05$ and $N_{\rm th}=0.14 \pm 0.04$)
and phase squeezing ($r = 0.46 \pm 0.04$ and $N_{\rm th}=0.16 \pm 0.03$)
for different values of the (real) amplitude $\alpha$.
The solid lines correspond to the theoretical previsions. The insets show
the experimental homodyne traces processed in order to obtain the value
of $g^{(2)}_{\rm sGs}(0)$ (marked with the corresponding capital letters)
according to the theory presented in section~\ref{s:zero:g2}.
}\label{f:g20:exp}
\end{figure}
\par
After collecting the homodyne traces for a given state we first 
checked the Gaussianity of the input state by assessing the 
Kurtosis of the data sample as well as a more comphrensive battery
of Gaussianity test \cite{buono:10,olom}.  Then, we evaluated
the needed expectations by using the pattern function tomography, 
which allows one to reconstruct 
the moments of a given quadratures upon exploiting 
the whole data sample, thus reducing the statistical errors.
The results are reported in Fig.~\ref{f:g20:exp}, where we show the experimentally obtained values of 
$g^{(2)}_{\rm sGs}(0)$ (dots) 
as a function of the displacement (real) amplitude $\alpha$
for both phase and amplitude squeezing together with the theoretical predictions (solid lines).
According to the theoretical results, 
in the presence of amplitude squeezing we find
a threshold on $\alpha$ above which $g^{(2)}_{\rm sGs}(0)<1$, whereas the phase
squeezed states always lead to a positive second-order correlation function. In the same
figure we also report the raw homodyne 
traces corresponding to some of the
experimental points: features of 
amplitude squeezing (A, B and C) or phase 
squeezing (D, E and F) are clearly seen.
\section{Conclusions}\label{s:concl}
In conclusion, we have suggested and demonstrated a 
reconstruction scheme for the zero-delay-time second-order 
correlation function $g^{(2)}(0)$ of Gaussian states
and we have proved it experimentally for single-mode states.
Our scheme is based on a single homodyne detector and the quantum state 
reconstruction of Gaussian state by pattern function tomography.
The results are based on the analytic expression of the correlation function 
$g^{(2)}(0)$ in terms of the elements of covariance matrix and the displacement
amplitude of the Gaussian state, which also show that
when the input state is nonclassical, there exists a threshold 
value of the coherent amplitude, and a range of values of the
complex squeezing parameter, above which $g^{(2)}(0) < 1$.
For amplitude squeezing and real coherent amplitude, the 
threshold is a necessary and sufficient condition for the nonclassicality of the state.
Our technique allows us to reliably characterise photon-number nonclassicality of
Gaussian states without the need of photon-resolving detectors.
Eventually, the recent achievements we obtained in the
experimental reconstruction of symmetric two-mode squeezed thermal states
\cite{cialdi:16} pave the way for further investigation of the second-order correlation
function beyond the single-mode case, which will be thoroughly addressed
in future works.
\section*{Acknowledgments}
This work has been supported by the University of Milan 
through the project CVQTIQP (grant 15-6-643), by JSPS 
through FY2017 program (grant S17118) and by SERB through the
VAJRA award (grant VJR/2017/000011).

\end{document}